\documentclass[%
reprint,
superscriptaddress,
amsmath,
amssymb,
aps,
]{revtex4-2}

\usepackage{CJK}
\usepackage{graphicx}
\usepackage{epstopdf}
\usepackage{dcolumn}
\usepackage{bm}

\usepackage{flushend}

\usepackage
[colorlinks,
linkcolor=blue,
anchorcolor=blue,
urlcolor=blue,
citecolor=blue
]{hyperref}

\usepackage{color}
\usepackage{upgreek}
\usepackage{makecell}

\usepackage{textcomp}

\usepackage{threeparttable}

\usepackage{gensymb}

\usepackage{hyperref}

\graphicspath{{figure/}} 
\usepackage{subfigure} 
\usepackage[utf8]{inputenc}
\usepackage[T1]{fontenc}
\usepackage{mathptmx}
\usepackage{etoolbox}
\DeclareUnicodeCharacter{202F}{\,}
\usepackage{CJK}
\usepackage{epstopdf}
\usepackage{float} 
\usepackage{siunitx} 
\usepackage[T1]{fontenc} 
\usepackage{textcomp} 
\usepackage{amsmath} 
\usepackage{color}
\usepackage{upgreek}
\usepackage{makecell}
\usepackage{textcomp}
\usepackage{threeparttable}
\usepackage{gensymb}
\usepackage{url}
\usepackage{booktabs}
\usepackage{tabularx}

\usepackage{mhchem}

\begin{document}
	
	\preprint{APS/123-QED}
	
	\title{Proton-CAT: a Novel Strategy for Enhanced Proton Therapy}
	
	\author{Zhao Sun}
	\thanks{These authors contributed equally to this work.}
	\affiliation{Key Laboratory of Radiation Physics and Technology of Ministry of Education, Institute of Nuclear Science and Technology, Sichuan University, Chengdu 610064, China}
	
	\author{Zhencen He}
	\thanks{These authors contributed equally to this work.}
	\affiliation{West China School of Basic Medical Sciences and Forensic Medicine, Sichuan University, Chengdu 610041, China}
	
	\affiliation{Key Laboratory of Radiation Physics and Technology of Ministry of Education, Institute of Nuclear Science and Technology, Sichuan University, Chengdu 610064, China}
	
	\author{Zhuohang He}
	\affiliation{Key Laboratory of Radiation Physics and Technology of Ministry of Education, Institute of Nuclear Science and Technology, Sichuan University, Chengdu 610064, China}
	
	\author{Junxiang Wu}
	\affiliation{Key Laboratory of Radiation Physics and Technology of Ministry of Education, Institute of Nuclear Science and Technology, Sichuan University, Chengdu 610064, China}
	\affiliation{Radiation Oncology Department Sichuan Cancer Hospital and Institute, Affiliated Cancer Hospital of University of Electronic Science and Technology of China, Chengdu 610041, China}
	
	\author{Liyuan Deng}
	\affiliation{Key Laboratory of Radiation Physics and Technology of Ministry of Education, Institute of Nuclear Science and Technology, Sichuan University, Chengdu 610064, China}
	
	\author{Ziqi Chen}
	\affiliation{Key Laboratory of Radiation Physics and Technology of Ministry of Education, Institute of Nuclear Science and Technology, Sichuan University, Chengdu 610064, China}

	\author{Junkang Jiang}
	\affiliation{Key Laboratory of Radiation Physics and Technology of Ministry of Education, Institute of Nuclear Science and Technology, Sichuan University, Chengdu 610064, China}
	
	\author{Hang Zhu}
	\affiliation{Key Laboratory of Radiation Physics and Technology of Ministry of Education, Institute of Nuclear Science and Technology, Sichuan University, Chengdu 610064, China}
	
	\author{Bingwen Zou}
	\affiliation{Department of Thoracic Oncology, Cancer Center, West China Hospital, Medical School, Sichuan University, Chengdu, 610041, China}
	
	\author{Shuyu Zhang}
	\email{zhang.shuyu@hotmail.com}
	\affiliation{West China School of Basic Medical Sciences and Forensic Medicine, Sichuan University, Chengdu 610041, China}
	
	\author{Zhimin Hu}
	\email{huzhimin@scu.edu.cn}
	\affiliation{Key Laboratory of Radiation Physics and Technology of Ministry of Education, Institute of Nuclear Science and Technology, Sichuan University, Chengdu 610064, China}
	
	\date{\today}
	
	\begin{abstract}
		
	We present a nitrogen-targeting-Proton-Carbon-Alpha-Therapy method, abbreviated as Proton-CAT. It achieves the partial conversion of protons to carbon ions ($^{12} \text{C}$) and $\alpha$ particles through nuclear reactions between protons and nitrogen-15 ($^{15} \text{N}$), thus coupling the benefits of carbon-ion therapy into the conventional proton therapy. Monte Carlo simulations validated the effectiveness of the Proton-CAT, and the study specifically focused on the distribution of relative energy deposition. The results indicated that the presence of $^{15} \text{N}$ enhanced the maximum dose level of protons, resulting in more effective damage confined to tumor cells. Statistical analysis of secondary ions has shown that the Proton-CAT significantly increases the production efficiencies of $^{12} \text{C}$ and $\alpha$ particles. Furthermore, it has been revealed that elevating the $^{15} \text{N}$ concentration significantly boosts the dose of $^{12} \text{C}$ and $\alpha$ particles within the tumor region. The present work would  contribute to the future development of proton therapy.
		
	\end{abstract}
	
	\maketitle
	
	
	
	In this century, cancer has been the leading cause of premature death globally and the greatest obstacle to further increasing life expectancy \cite{ soerjomataram2021planning}. Radiation therapy is an effective clinical treatment for malignant tumors. Currently, most clinical radiation therapy uses photon beams (e.g., X- and $\gamma$-rays) or electron beams with energies ranging from several MeV to tens of MeV \cite{park2012photon, hogstrom2006review}. If the dose is high enough, radiation beams can destroy any cancer cells or living organisms. Considering the physical properties of photons, normal tissue around the target volume still receive a momentous and unnecessary dose of radiation. Thus, the critical factor in the development of radiation medicine has been how to avoid irradiating normal tissue. 
	
	For many years in parallel, proton and heavy ion radiation therapies have been developed tempestuously \cite{durante2017charged, PhysRevA.105.062822, liamsuwan2014microdosimetry, frese2012mechanism}. They utilize the physical characteristics of the Bragg peak formation by protons or heavy ions traversing through matter. When protons and heavy ions slow down by transferring momentum, the stopping power near the Bragg peak velocity significantly increases \cite{shepard2023electronic}. Therefore, the dose distribution of protons and heavy ions is better than that of photons. Adjusting the high-dose zone embedded in the tumors, it is expected to increase the tumor dose and minimize the damage to normal tissue. Protons and heavy ions interacting with tissue and organisms through ionization can result in localized energy deposition. This process results in the creation of highly energetic holes on the DNA chains, which serve as a source of oxidative damages, and thus causes biological effects such as mutations, chromosomal aberrations, and cell death \cite{shepard2023electronic, moding2013strategies}. 
	
	However, protons primarily induce cell death indirectly by causing single-strand breaks (SSBs). Cancer cells would be effectively killed if the next SSB could cut another strand before the damaged DNA recovers, otherwise they would be repaired \cite{souici2017single, schardt2010heavy}. It is difficult for protons to effectively treat chemoresistant and hypoxic tumors \cite{schardt2010heavy}. For this reason, researchers widely regard carbon ions as the optimal heavy ions for therapeutic purposes. Carbon ions possess the Bragg peak which is similar to protons. Carbon ions exhibit higher linear energy transfer (LET) values and energy deposition density per unit of energy than protons. The relative biological effectiveness (RBE) of carbon ions is 2 to 3 times higher than that of protons \cite{liamsuwan2014microdosimetry, frese2012mechanism}. The reason is that the carbon ions have the ability to directly kill tumor cells by inducing double-strand breaks (DSBs), which makes the damaged cells more difficult to be repaired \cite{souici2017single, solov2009physics}.  Furthermore,  due to the hypoxic nature of cancer cells, carbon ions exhibit a lower oxygen enhancement ratio (OER, the ratio of radiation doses required to cause the same damages in hypoxic and oxygenated cells) compared to that of protons \cite{suit2010proton, wenzl2011modelling}. Higher RBE and lower OER values are characteristic biological advantages of the carbon ions with high LET. However, carbon-ions therapy demands higher requirements for equipment and technology, making it more expensive than the proton therapy, which also restricts its widespread clinical adoption \cite{nunes2015protontherapy}.
	\begin{figure*}
		\centering
		\setlength{\abovecaptionskip}{0.5cm}
		\includegraphics[width=0.9\textwidth]{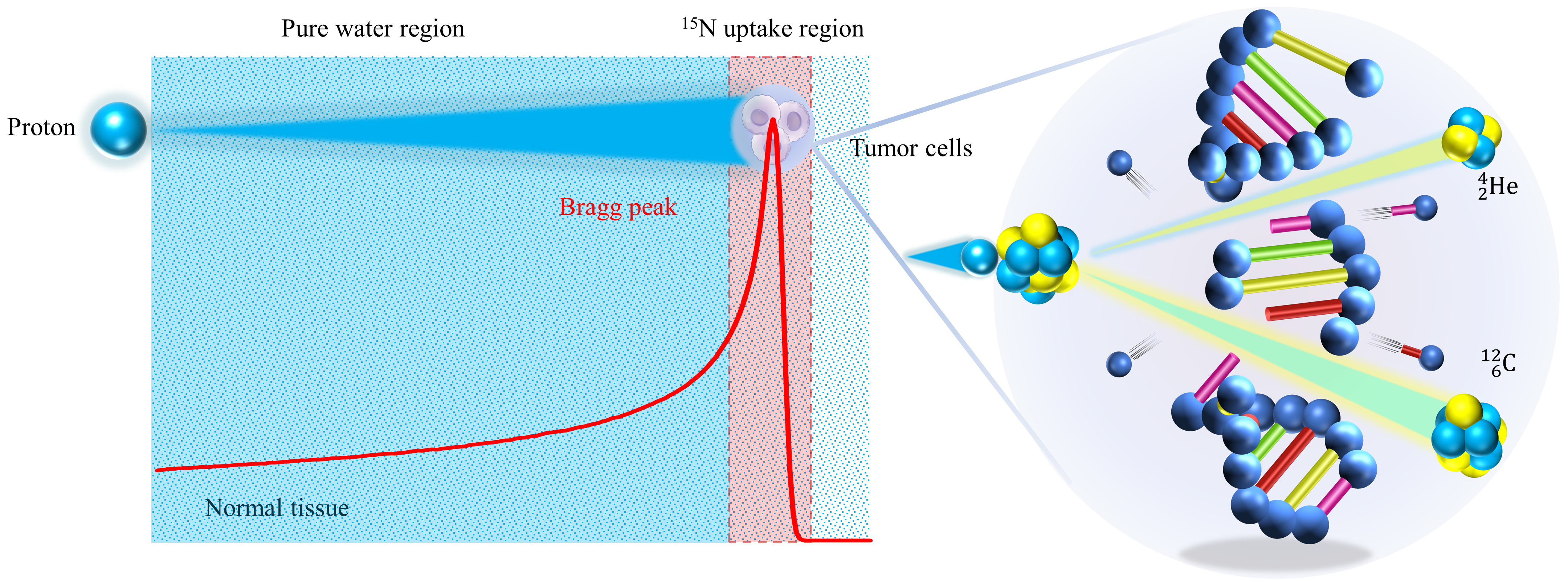}
		\vspace{-6mm} 
		\caption{\label{fig1}Conceptual diagram of the Proton-CAT and principles of geometric construction in Monte Carlo simulations. In the simulation, $^{15} \text{N}$-rich tumor phantom of 0.5 cm thickness covered the Bragg peak of the 75 MeV proton beams, and the tissue thickness in front of the tumor target was 4.35 cm. }
		\vspace{-4.5mm}
	\end{figure*}
	
	In this Letter, we present a novel proton therapy method, termed nitrogen-targeting-Proton-Carbon-Alpha-Therapy method (Proton-CAT), which leverages the advantages of protons and carbon ions. By using Monte Carlo simulations, we validated the efficacy of the Proton-CAT method. The simulation results demonstrated that the Proton-CAT can generate significant secondary particles which are carbon-12 ($^{12} \text{C}$) ions and $\alpha$ particles. Moreover, compared with the conventional proton therapy, these secondary particles can enhance the absorbed dose several times in tumor regions. The Proton-CAT harnesses the inherent benefits of proton therapy and amplifies its efficacy by incorporating the therapeutic potential of $^{12} \text{C}$ and $\alpha$, offering a promising avenue for advanced cancer treatment strategies. 
	
	The conceptual diagram of the Proton-CAT is shown in Fig. \ref{fig1}. It uses the proton and stable nitrogen isotope ($^{15} \text{N}$) fusion reaction to convert protons into $^{12} \text{C}$ and $\alpha$:
	$$ p+^{15}\text{N} \rightarrow ^{16} \text{O}^* \rightarrow ^{12} \text{C}+ \alpha +  \text{5.0 MeV}. $$ 
	This nuclear reaction has a reasonable cross section when the proton energy is between 3 and 11 MeV, which produces large numbers of secondary particles and releases large amounts of energy \cite{schardt1952disintegration,bashkin1959cross, nurmukhanbetova2017implementation}. Simultaneously, the interaction between protons and $^{15} \text{N}$ leads to the fragmentation of the excited oxygen ($^{16} \text{O}$) into $^{12} \text{C}$ and $\alpha$. As shown in Fig. \ref{fig1}, specialized tumor-specific drugs enriched with $^{15} \text{N}$ dose-enhancement agent (DEA) are administered to the tumor region in practical therapy. When irradiated with the external proton beam, part of the protons are converted into $^{12} \text{C}$ and $\alpha$ particles via Proton-CAT in the $^{15} \text{N}$-uptake region. These secondary short-range radiation sources are expected to cause more DSBs to the DNA of tumor cells while minimizing cellular damage to normal tissue. Finally, proton therapy has acquired the advantages of heavy ion therapy. This approach, where each absorbed unit of energy causes additional damage to tumor cells, holds the potential to fundamentally enhance the efficiency of radiation therapy \cite{Lipengolts_2021, DURANTE2018160}. In clinical settings, tumor cells e.g., gastric, colorectal, and breast cancers have higher glutamine ($\text{C}_5 \text{H}_{10} \text{N}_2 \text{O}_3$) uptake than normal cells \cite{reinfeld2021cell}. Therefore, $^{15} \text{N}$-labeled glutamine can be applied to the Pronton-CAT as DEA that meets the requirements for high tumor uptake, low normal tissue uptake, low toxicity and rapid clearance after treatment \cite{barth2018realistic}. Compared to normal cells, if the DEA uptake in the tumor was high enough, then the curing effect was achieved. This characteristic provides the possibility for clinical application of the Proton-CAT.
	
	Herein, we further investigated the potential benefits of enhancing $^{15} \text{N}$ concentration in the Proton-CAT for proton therapy. Additionally, the secondary particles produced by Proton-CAT were simulated, aiming to quantify the contribution of each nuclear reaction to secondary particles and assess whether they could achieve a synergistic effect for dose enhancement. 
	
	The Geant4 toolkit was used for the simulations, an open-source Monte Carlo simulation program widely used in medical physics \cite{agostinelli2003geant4}. In the present work, the physical list employed for all simulations is G4HadronPhysics QGSP\_BIC\_AllHP. It employs the binary cascade de-excitation model, electromagnetic physics option4 as the standard, and a precision data-driven model for low-energy protons, neutrons, deuterium, tritium, $^{3} \text{He}$, and $\alpha$ particles. An assessment of both the binary cascade and standard electromagnetic physics option4 has been conducted to ensure optimal agreement between simulated results and measurements in proton therapy \cite{jarlskog2008physics}. In addition, it is created with the High-Precision Charged Particle physics model, which is suitable for energy below 200 MeV. The range threshold for all secondary production is set to 0.1 mm, under which the projectile trajectory is terminated and all remaining kinetic energy is deposited locally.
	
	\begin{table}[]
	\renewcommand{\arraystretch}{1.5}
	\caption{Compositions, mass density, and atomic proportion of the three-group tumor phantom.}
	\label{table1}
	\begin{tabular*}{\linewidth}{@{}ccccc@{}}
		\toprule[1.5pt]
		
		Groups        				& Mass density ($g/cm^3$) 	& \makebox[0.065\textwidth]{$^{1} _1 \text{H}$}    	& \makebox[0.065\textwidth]{$_{8}^{16}\text{O}$}   	& \makebox[0.065\textwidth][c]{$^{15} _7 \text{N}$} \\ 
		\toprule[1.0pt]
		Pure water    				& 1.000                		& 66.7 	& 33.3 	& 0.0   \\
		with 10\% $^{15} \text{N}$ 	& 1.139                		& 60.0  & 30.0  & 10.0  \\ 
		with 30\% $^{15} \text{N}$ 	& 1.538                		& 46.7 	& 23.3 	& 30.0  \\ 
		\bottomrule[1.5pt]
	\end{tabular*}
	\end{table}
	In the simulation, the geometry consisted of water and tumor phantom. As shown in the blue zone in Fig. \ref{fig1}, considering that the tissue is mainly composed of water, the material representing normal tissue in the simulation geometric model was defined as pure water. To reflect the dose enhancement brought about by the Proton-CAT, the material within the tumor phantom was defined as a mixture of water and varying concentrations of $^{15} \text{N}$ (red zone in Fig. \ref{fig1}). As shown in Table \ref{table1}, there were three uptake setups for the tumor region, namely, pure water ($100\% \ \text{H}_2\text{O}$), with 10\% $^{15} \text{N}$ ($90\% \  \text{H}_2\text{O} + 10\% \  ^{15} \text{N}$), and with 30\% \ $^{15}  \text{N}$ ($ 70\% \  \text{H}_2\text{O} + 30\% \  ^{15} \text{N}$). Both the normal tissue and tumor phantom were defined as cubes with a width of 5 cm. A proton beam with assigned energy was directed from the front of the water, covering the entire target volume. Therefore, the thickness of water and tumor models was set to 4.35 cm and 0.5 cm, respectively. 75 MeV incident energy was set to ensure coverage of the entire Bragg peak over the tumor region and the center of the tumor region corresponds to the point of maximum dose delivery for each proton. For each simulation run, a total of $10^9$ protons were transported through the target volume.

	\begin{figure}
		\centering
		\setlength{\abovecaptionskip}{0.5cm}
		\includegraphics[width=0.48\textwidth]{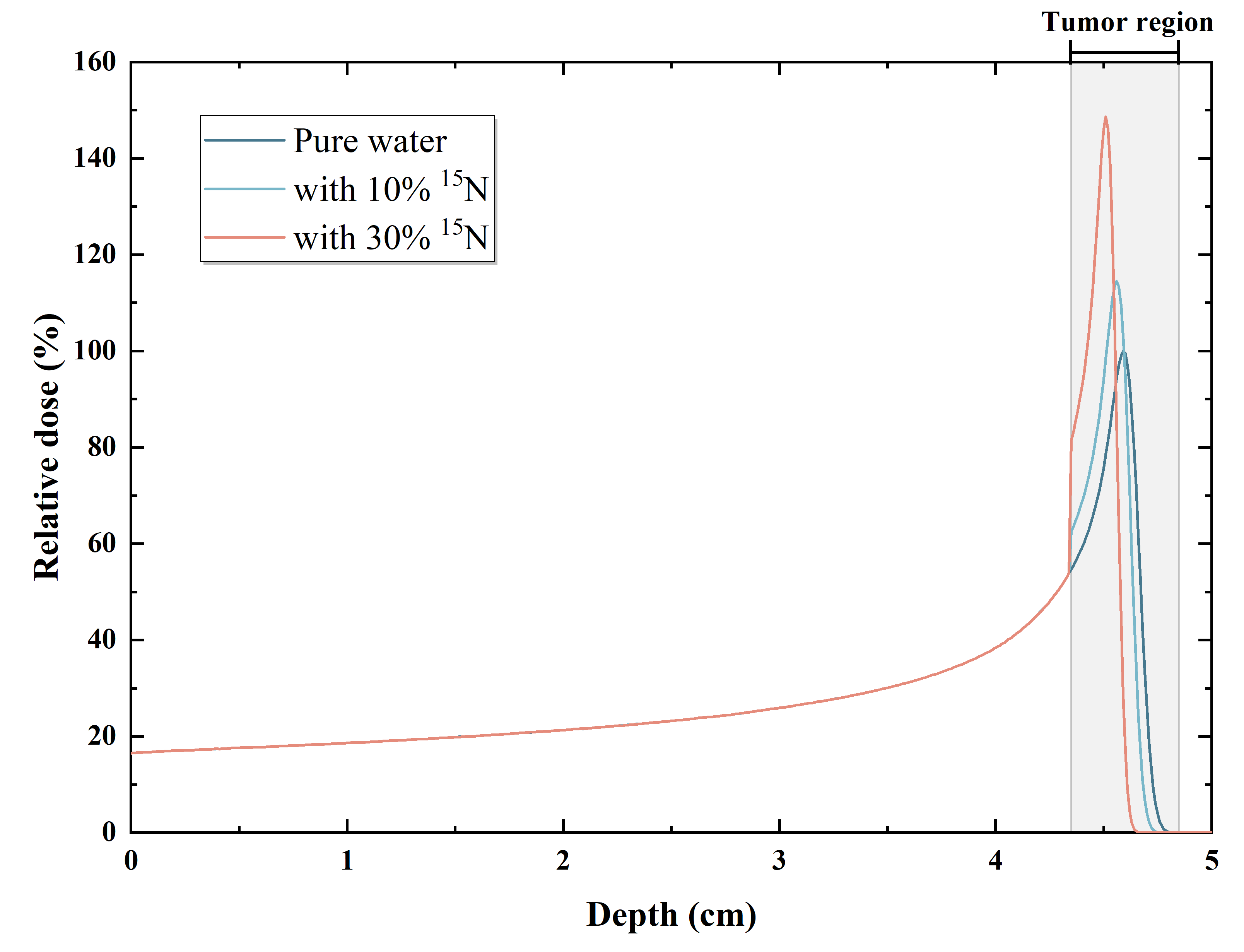}
		\vspace{-6mm} 
		\caption{\label{fig2} PDD of protons in three different material settings when the point of maximum dose is located in the tumor region: the black line is the normalized PDD from the pure water phantom, while the blue and red lines represent 10\% and 30\% $^{15} \text{N}$ doping, respectively.}
		\vspace{-4.5mm}
	\end{figure}
	Fig. \ref{fig2} shows the percentage dose depth (PDD) in the target volume of the proton. The PDD in pure water was used as the reference, increasing with depth, and reaching a Bragg peak near the maximum depth. Compared to pure water, doping with $^{15} \text{N}$ can enhance the dose in tumor volume and cause a forward shift of the Bragg peak. As the protons released their kinetic energy of the last few MeV (the energy range with the largest nuclear reaction cross-section) when they reached the tumor region, this increased the energy deposition of the Proton-CAT \cite{mazzone2019effectiveness}. These secondary particles produced from the nuclear reaction between protons and $^{15} \text{N}$ are responsible for a significant portion of dose delivery, and they deposit energy primarily through the ionization processes. These deliveries become more considerable as the concentration of $^{15} \text{N}$ increases to 30\%. In addition, as shown in Fig. \ref{fig2}, it can be seen that the accumulation of range straggling tends to broaden the peak with increasing depth. Nonetheless, the monoenergetic proton beam has a finite range and the deposited energy is not uniform in the tumor volume. Moreover, this finite region becomes more constrained with the variations of doping concentration within the tumor region.

	\begin{figure}
		\centering
		\setlength{\abovecaptionskip}{0.5cm}
		\includegraphics[width=0.48\textwidth]{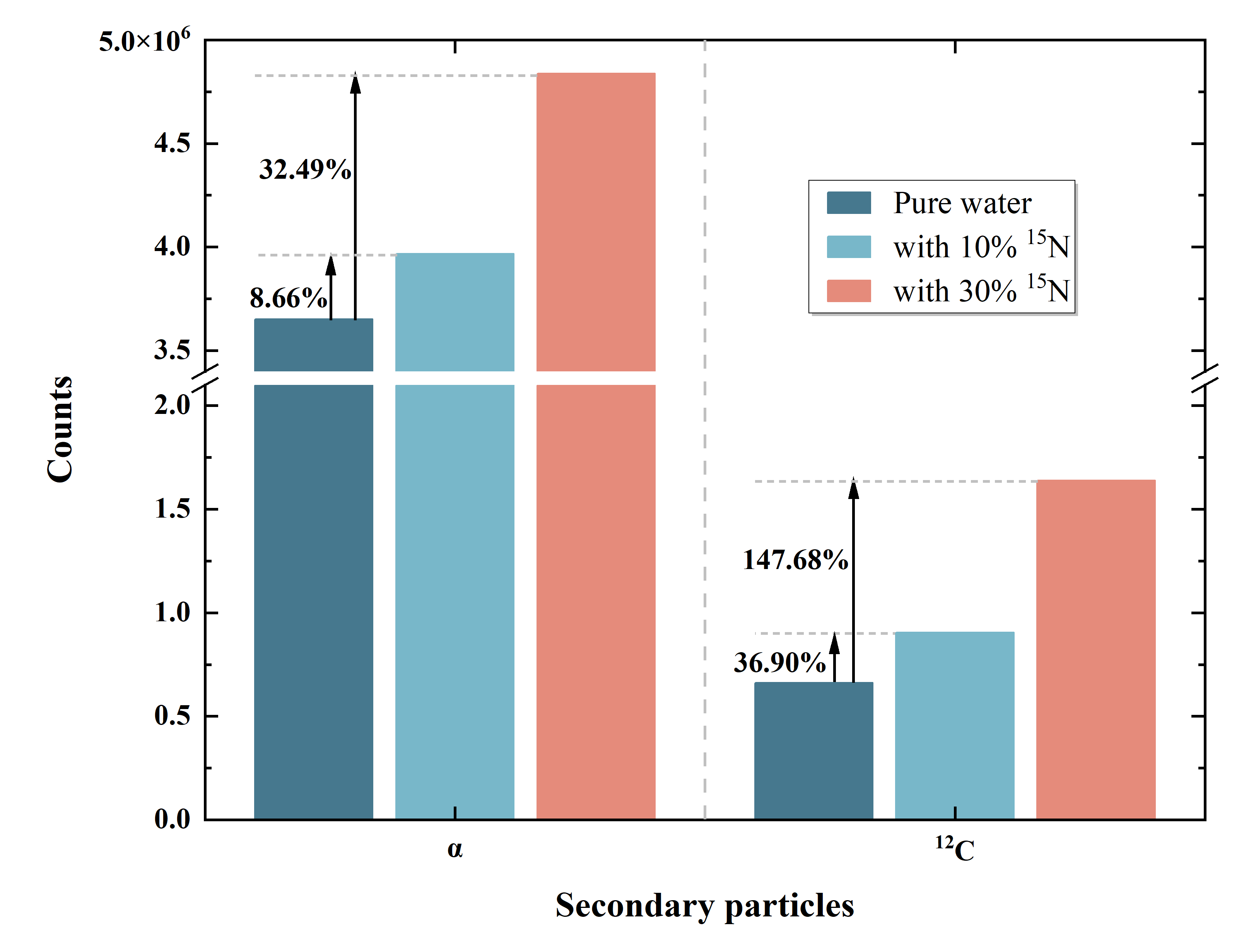}
		\vspace{-6mm} 
		\caption{\label{fig3}The counts of secondary particles generated by protons in the tumor region compared to a pure water phantom, as well as the gain relative to normal tissue.}
		\vspace{-4.5mm}
	\end{figure}	
	
	This study exploits the nuclear reactions between protons and targeting elements to enhance the dose by producing heavy secondary particles. We counted the secondary particles (mainly $\alpha$ and $^{12} \text{C}$) produced in the tumor region, which are shown in Fig. \ref{fig3}. It can be seen that compared with the case of pure water, i.e., normal tissue, introducing $^{15} \text{N}$ into the tumor region will produce more $\alpha$ and $^{12} \text{C}$ particles. Moreover, with the introduction of $^{15} \text{N}$, the $^{12} \text{C}$ gain percentage was much greater than $\alpha$.  The gain is more pronounced as the $^{15} \text{N}$ concentration increases, i.e., more than a threefold increase in product particles results from a threefold increase in $^{15} \text{N}$ concentration. As mentioned before, these energetic secondary particles contribute to dose enhancement. Morever, compared to conventional proton therapy, $\alpha$ and $^{12} \text{C}$ particles with higher LET values can result in greater RBE. However,  for normal tissue (i.e., without $^{15} \text{N}$), $\alpha$ and $^{12} \text{C}$ particles will also be produced, attributed to nuclear reactions induced by the interaction between protons and $^{16} \text{O}$. The primary nuclear reactions generating $\alpha$ and $^{12} \text{C}$ particles in pure water are the $^{16} \text{O}(p, \alpha) ^{13}\text{N}$ and $^{16} \text{O}(p, p'\alpha) ^{12}\text{C}$ nuclear reactions, with relatively low thresholds of 5.66 MeV and 7.16 MeV, respectively \cite{cho2017feasibility, chapman1967proton, chapman1967mechanism}. It should be noted that for tumors containing 10\% and 30\% concentrations of $^{15} \text{N}$, the interaction of protons with $^{16} \text{O}$ in the water is another significant mechanism for generating alpha particles. Therefore, the number of $\alpha$ produced in the tumor region was greater than that of $^{12} \text{C}$. 
	
	\begin{figure*}
		\centering
		
		\setlength{\abovecaptionskip}{0.5cm}
		\subfigure{
			\label{fig4ab}
			\includegraphics[width=0.5\textwidth]{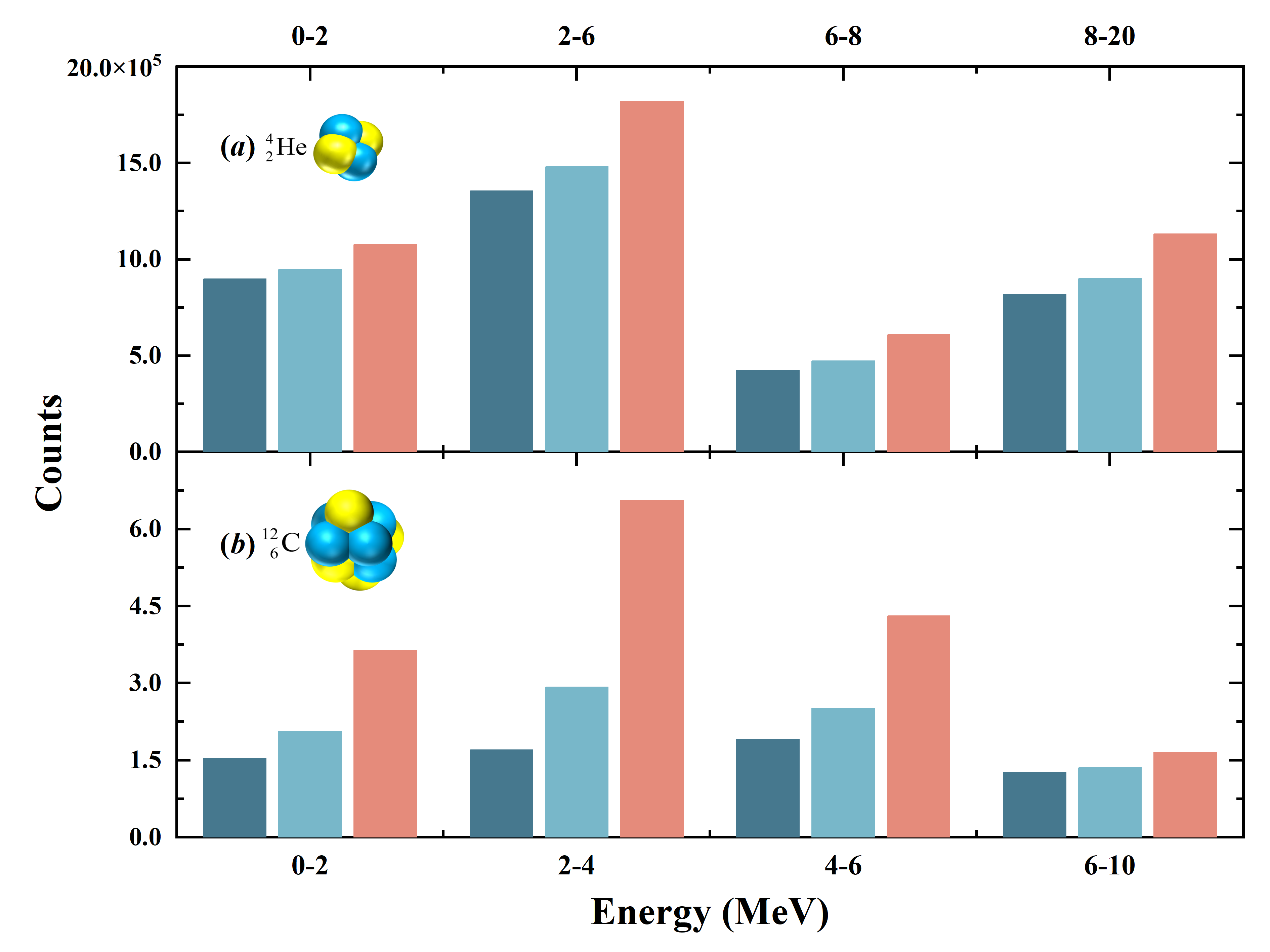}
			\hspace{-4mm}
		}		
		\vspace{-0mm}
		\subfigure{
			\label{fig4b}
			\includegraphics[width=0.49\textwidth]{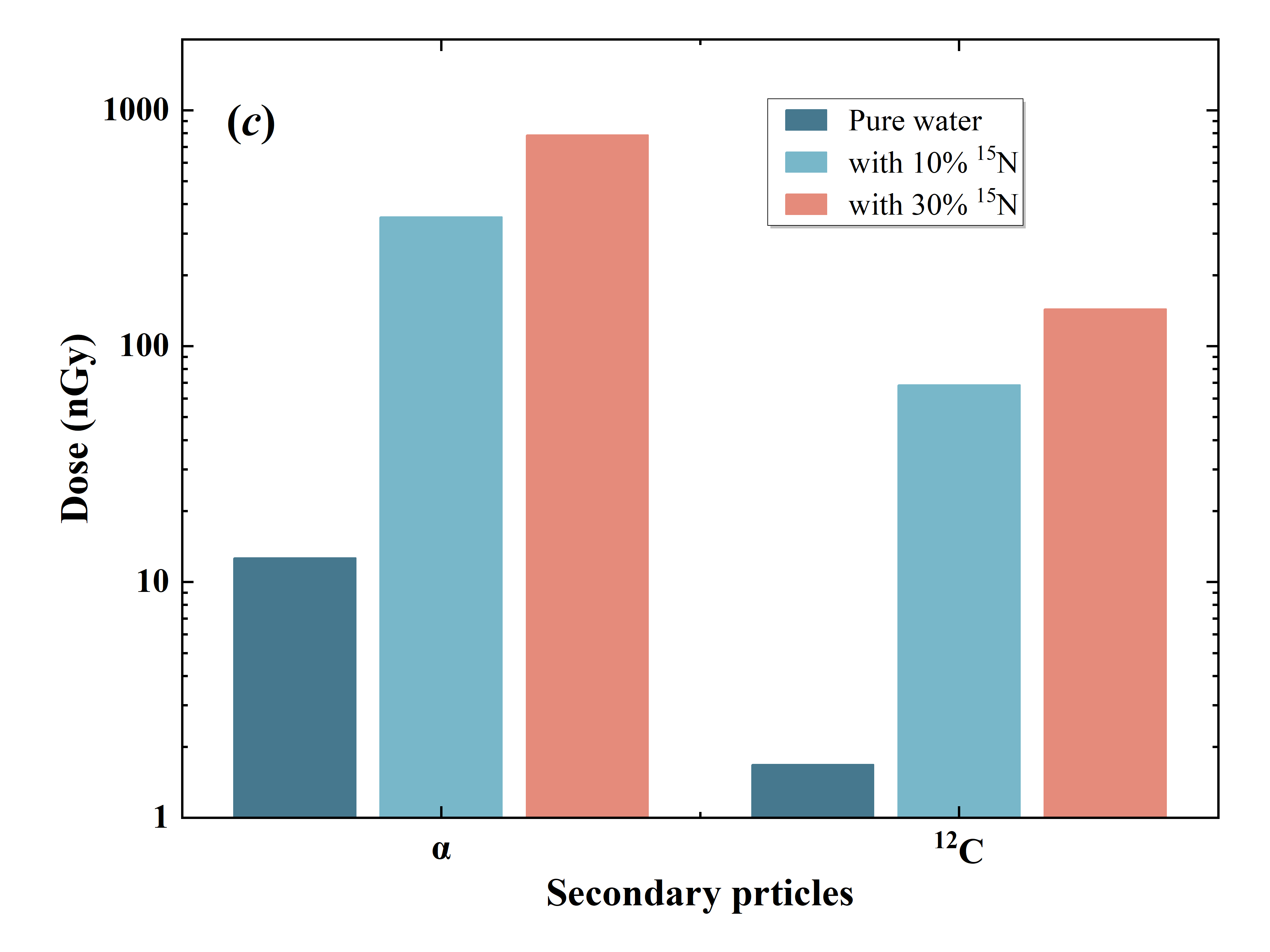}
			\hspace{-4mm}
		}
		
		\caption{\label{fig4} Left: Piecewise statistical results of ($a$) $\alpha$  and ($b$) $^{12} \text{C}$ particle energy generated in tumor region after $10^9$ protons are implanted. Right: The absorbed dose generated by proton-induced secondary particles ($\alpha$ and $^{12} \text{C}$) in the tumor region.}
		\vspace{-4mm}
	\end{figure*}
	
	As shown in Fig. \ref{fig4}, we performed a piecewise statistical analysis on the energy generated by $\alpha$ and $^{12} \text{C}$ particles in the tumor region. It can be seen that 75\% of the $\alpha$ energies are in the rang of $0-8$ MeV and more than 80\% of the $^{12} \text{C}$ particles are located in the rang of $0-6$ MeV. Furthermore, with increasing $^{15} \text{N}$ concentration, the main energy of the $^{12} \text{C}$ contribution from the Proton-CAT is located in the range of $2-4$ MeV. In the tumor region, the nuclear reaction of the Proton-CAT released the energy mainly in generating the secondary particles. Subsequently, the $\alpha$ and $^{12} \text{C}$ gradually slow down and LET increases sharply. In the case of $^{12} \text{C}$, its mass limits deviation, resulting in an almost rectilinear trajectory, with a corresponding penetration range of less than 10 \si{\micro\metre}, which is sufficient to damage the tumor nucleus with a normal size. On the other hand, the energy range of the emitted $\alpha$ particles is similar to that used in radiopharmaceuticals. As a result, the penetration range of $\alpha$ is confined mostly within 100 \si{\micro\metre} and can damage cells in the vicinity of the reaction site \cite{james2021current}. However, it is worth noting that nuclear reactions between protons and $^{16} \text{O}$ also generate small amounts of $\alpha$ and $^{12} \text{C}$. These secondary particles are distributed over all energy ranges, leading to higher dose deposition in the tumor region.
	
	From the results above, it can be seen that the Proton-CAT increases the deposited energy in the tumor region. The therapeutic effect is attributed to both the primary and secondary particles. As shown in Fig. \ref{fig2}, the Proton-CAT generates a large number of secondary particles, which contribute to the total absorbed dose. Their dose deposition contributes to the RBE of the proton beam in the tumor region at the same location as the formation of the new Bragg peak \cite{schardt2010heavy}. Therefore, we also calculated the dose increase caused by the secondary particles within the tumor region, as shown in Fig. \ref{fig4} (c). It can be seen that when 75 MeV protons were implanted, the absorbed dose of the secondary particles produced by nuclear reactions increased significantly compared to the pure water. The dose produced by secondary particles was much higher than that delivered without adding $^{15} \text{N}$. Similarly, the Proton-CAT exhibits an expected increase in absorbed dose with the increase of $^{15} \text{N}$ concentration. As shown in Fig. \ref{fig4} (b), the Proton-CAT can generate numerous $^{12} \text{C}$ with $0-6$ MeV, further enhancing the dose within the tumor region. This is the greatest advantage of the Proton-CAT. Moreover, the total absorbed dose deposition of secondary particles can be highly dependent on the circumstances, including the incident proton energy, the geometric size of the tumor, and the concentration of $^{15} \text{N}$.
	
	In conclusion, a novel enhanced proton therapy approach, Proton-CAT, was proposed based on the nuclear reaction between protons and $^{15} \text{N}$. Monte Carlo simulations were employed to evaluate the performance of the Proton-CAT in proton therapy. In the simulations, different concentrations of $^{15} \text{N}$ were set to compare the effects of concentration enhancement on the dose deposition and the generation of secondary particles. Their average dose in the tumor region was calculated to assess the dose deposition efficacy of secondary particles in various scenarios. These results indicated that the Proton-CAT demonstrates a dose enhancement advantage in proton therapy, with the secondary particle $^{12} \text{C}$ playing a crucial role therein. It must be emphasized that although our results confirm the essential efficacy and advantages, further validation is required for the clinical application of this method. Future research will include in vitro experiments based on the current simulation calculations as the next step for the clinical application of the Proton-CAT. 
	
	\vspace{3mm}
	This work was supported by the Sichuan Science and Technology Program (Grant Nos. 2023ZYD0017 and 2022NSFSC0797), the Defense Industrial Technology Development Program (Grant No. JCKYS2023212808), the National Natural Science Foundation of China (Nos. 12374234 and 12074352), and the Fundamental Research Funds for the Central Universities in China (Grant No. YJ202144).

	\bibliography{ref}

\end{document}